# Translation of Nanoantenna Hot-Spots by a Metal-Dielectric Composite Superlens

**Zhengtong Liu (刘政通)\*, Mark D. Thoreson, Alexander V. Kildishev, and Vladimir M. Shalaev**

*School of Electrical and Computer Engineering and Birck Nanotechnology Center, Purdue University,*

*West Lafayette, IN 47907, USA*

**Abstract**: We employ numerical simulations to show that highly localized, enhanced electromagnetic fields, also known as "hot spots," produced by a periodic array of silver nanoantennas can be spatially translated to the other side of a metal-dielectric composite superlens. The proposed translation of the hot spots enables surface-enhanced optical spectroscopy without the undesirable contact of molecules with metal, and thus it broadens and reinforces the potential applications of sensing based on field-enhanced fluorescence and surface-enhanced Raman scattering.

Optical nanoantennas have been of great interest recently due to their potential applications in biosensors, near-field scanning optical microscopy (NSOM), quantum optical information processing, enhanced Raman scattering and other optical processes [1-6]. A nanoantenna made of paired metal nanoparticles can support a highly efficient, localized surface plasmon resonance and produce a significantly enhanced and highly confined electromagnetic field [7-9], which is also known as a "hot spot" [10-14]. These hot spots are especially useful in sensing applications such as surface-enhanced Raman scattering (SERS) [15] and enhanced fluorescence [16-21]. However it has been reported that the fluorescence and SERS enhancement depends on the separation between the fluorophores and the metal nanoparticles, and if the fluorophores are too close to the nanoparticles, the fluorescence may be quenched instead of enhanced [22-23]. Another problem is that when in direct contact with metal surfaces, some molecules experience significant structural (denaturation) and/or functional changes [24-26]. A metal-dielectric composite lens offers a solution to these problems by "imaging" the high local fields produced by a SERS or fluorescence substrate to the other side of the composite lens, where bio-molecules or other analytes are

Email: liuz@purdue.edu

placed. Here we report on simulations in which we consider the translation of nanoantenna-enhanced fields through a metal-dielectric near-field superlens.

Research on superlensing has rapidly progressed in a short few years. A planar material slab with simultaneously negative permittivity and permeability can focus propagating waves and enhance evanescent waves, thereby acting as a perfect lens or superlens that does not suffer from the diffraction limit [27]. The operation of a perfect lens requires that $\varepsilon_1 = -\varepsilon_2$ and $\mu_1 = -\mu_2$, where $\varepsilon_1$ and $\mu_1$ are the permittivity and permeability of the perfect lens, and $\varepsilon_2$ and $\mu_2$ are those of the host material. If the source object is close to the slab, the electrostatic approximation is valid and the superlensing requirement can be relaxed to $\varepsilon_1 = -\varepsilon_2$ [27]. This condition can be readily satisfied by metals such as gold and silver at particular frequencies. Such a thin metal slab is often referred to as a near-field superlens (NFSL) [27-28]. If a resonant nanoantenna is placed close to the NFSL, the lens can generate the image of the antenna's hot spot on the other side of the lens. Near-field superlenses based on bulk metal can operate only at a single frequency, determined by the material properties of the bulk host and metal. For the best plasmonic material, silver, this operational frequency turns out to be in the near ultraviolet. In contrast, the operational frequency for a metal-dielectric composite superlens can be varied through the visible and near infrared ranges by simply changing the metal filling factor [29]. Therefore, rather than limiting our analysis to bulk metal lenses, we obtain the NFSL operational condition by using a metal-dielectric composite [29]. In this letter we show by numerical simulations that a nanoantenna-superlens device can produce translated hot spots on the "image" side of the lens with somewhat decreased but still enhanced amplitudes.

The unit cell of our nanoantenna array combined with a superlens is illustrated in Fig. 1. Each unit cell consists of a nanoantenna composed of two closely spaced elliptical silver cylinders and a slab of composite silver-silica NFSL material below the nanoantenna. Both the nanoantenna and the NFSL slab are embedded in an infinite region of silicon host material. The structural dimensions are as follows: major axis 100 nm, minor axis 50 nm, gap 10 nm, antenna thickness 20 nm, superlens thickness 40 nm.

The separation between the antenna and the superlens is 20 nm. Light is normally incident on the device from the nanoantenna side of the superlens. In our discussions that follow, the superlens interface closer to the nanoantenna is referred to as the top interface, while the other side is the bottom interface.

Nanoantenna arrays have been studied experimentally and numerically in our previous work, where it has been shown that the arrays can be accurately modeled using the finite element method (FEM) [30]. The silver-silica composite superlens is modeled with an effective permittivity calculated using effective medium theory (EMT) [29]. For our analysis, the NFSL volume filling factor was set at 0.71 and the dimensionality was set at 2 [29]. Because of the relatively large permittivity of the composite, we used silicon as the adjacent dielectric host. The use of a dielectric host with a large permittivity also allows one to move the operational wavelength of the composite NFSL outside the strong absorption band [29]. The dielectric constants of the composite and silicon are shown in Fig. 2. At a wavelength of 1100 nm, where $\varepsilon_{EMT} = -20.18 + 0.336i$ and $\varepsilon_{Si} = 12.55 + 3 \times 10^{-4} i$, the requirement for superlensing operation is roughly satisfied. We note that the superlensing requirement cannot be completely satisfied because the composite has a relatively large imaginary part of dielectric permittivity, which cannot be matched by the dielectric constant of the host material (Si). The imaginary part of the composite's effective permittivity causes energy loss in the superlens and reduces the amplitude of translated fields. Therefore, a small imaginary part is preferred in order to obtain good results, and at 1100 nm the composite has the lowest imaginary part of effective permittivity. A commercial FEM software package (COMSOL Multiphysics) was used for our numerical simulations. The electric field-intensity enhancement (FE) along the center line of the unit cell is plotted in Fig. 3(a). The superlens material is located between $z = 0$ nm and $z = -40$ nm, while the nanoantenna elements are between $z = 20$ nm and $z = 40$ nm. At the bottom interface, the simulations show clearly that the FE peak occurs at an incident wavelength of 1100 nm. The FE along the center line at 1100 nm is plotted separately in Fig. 3(c). At the bottom interface, the FE is around 226. The same nanoantenna array without the superlens was also simulated; the FE in this case is plotted in Fig. 3(b) for comparison. In this case, the nanoantenna produces a hot spot at a wavelength of about 1160 nm.

However, the hot spot in this case is confined inside the gap, and the enhanced fields vanish completely only a few nanometers away from the gap, which is evident in Fig. 3(d). Comparing these two cases (with and without the superlens), we conclude that the field enhancement at the bottom interface is indeed the result of the superlensing effect, which enables the translation of enhanced hot spot fields from the top surface to the bottom surface. Note that the presence of the superlens shifts the resonant wavelength of the nanoantenna from 1160 nm to 1100 nm.

The electric field intensity enhancement map at the bottom interface of a unit cell is plotted in Fig. 4(a). The blue ellipses depict the outline of the nanoantenna pair. We observed that although the magnitude of the translated field enhancement is not as high as that inside the antenna gap without the superlens, the volume of the translated hot spot is much larger than the nanoantenna gap, which is beneficial for molecular detection techniques such as surface-enhanced Raman scattering. We calculated the enhancement at a distance of 20 nm below the bottom interface in order to understand the spatial extent of the translated fields; the results are plotted in Fig 4(b). Below the composite NFSL, the evanescent waves decay away from the superlens interface, and the FE drops to 48 at a distance 20 nm below the bottom interface (down from a value of 226 at the superlens bottom interface). However, the hot spot region is still present 20 nm from the bottom interface. The field enhancement of the same nanoantenna array embedded inside silicon without a superlens is plotted in Fig. 4(c) for comparison. In this case, no hot spot was observed at such a distance from the nanoantenna.

We have shown that by using a thin silver-silica composite layer as a near-field superlens, we can translate the hot spots generated by silver nanoantenna arrays to the other side of the superlens and thus move the hot spots away from the interface with metal. In practice, we expect that the geometry of the nanoantenna should be designed for the desired operational wavelength, while the silver volume filling factor of the silver-silica composite should be tailored to ensure that the superlens condition is satisfied. To achieve optimal conditions for fluorescence enhancement or Raman scattering measurements, the distance between the superlens and the molecules under detection can also be tuned using a thin dielectric cover layer at the bottom interface of the superlens. The proposed translation of nanoantenna hot spots to the other side of the superlens enables surface-enhanced optical spectroscopy without the unwanted side effects arising from contact between the molecules under study and the field-enhancing metal constituent.

The authors thank Vladimir D. Drachev for helpful discussions. This work was supported by ARO-STTR W911NF-07-C-0008.

**Figures:**

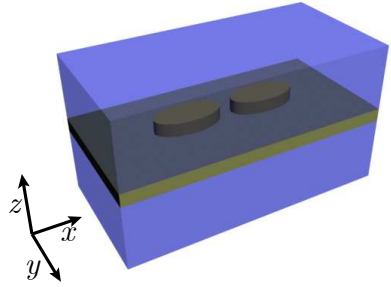

FIG 1: A unit cell of the nanoantenna array with superlens. The host material is silicon.

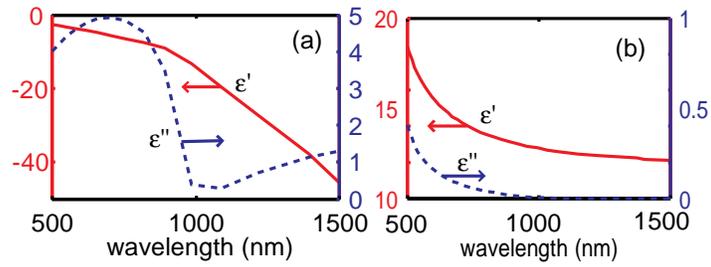

FIG 2: (a) The effective permittivity of silver-SiO2 composite calculated by EMT; (b) the permittivity of silicon.

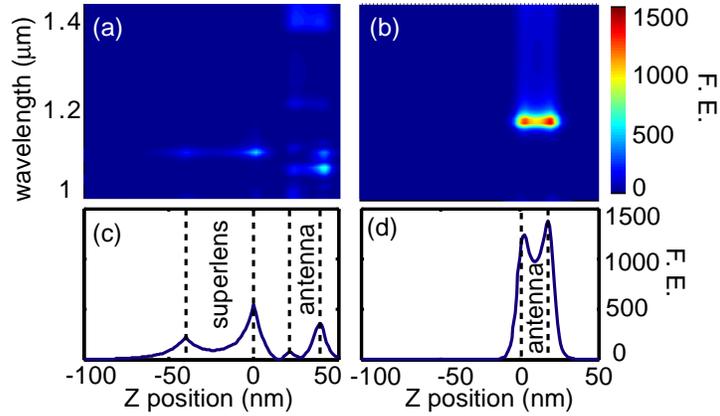

FIG 3: (a) The FE along the center line of a unit cell; (b) the FE along the center line of a nanoantenna without the superlens; (c) the FE along the center line at a wavelength of 1100 nm with the superlens; (d) the FE along the center line at a wavelength of 1160 nm without the superlens.

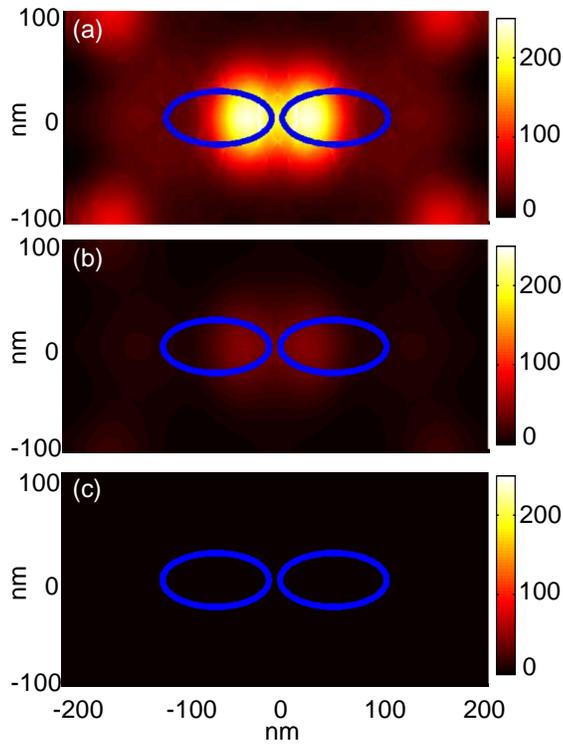

FIG 4: (a) The electric field intensity enhancement at the bottom interface and (b) 20 nm below the bottom interface of the superlens. Panel (c) shows the electric field intensity enhancement at the same distance as in (b) without the superlens (for comparison).